\documentclass{elsart}

\usepackage{natbib}

\usepackage{graphicx}
 
\begin{document}


\runauthor{Garrett, Porcas, Pedlar, Muxlow, \& Garrington}


\begin{frontmatter} 
\title{Wide-field VLBI Imaging} 
\author[JIVE]{M.A. Garrett}
\author[MPIfR]{R.W. Porcas}
\author[NRAL]{A. Pedlar} 
\author[NRAL]{T.W.B. Muxlow}
\author[NRAL]{S.T. Garrington}

\address[JIVE]{Joint Institute for VLBI in Europe, Postbus 2, 
7990~AA Dwingeloo, The Netherlands} 
\address[MPIfR]{Max-Planck-Institut f\"ur Radioastronomie,
Auf dem H\"ugel 69, Bonn 53121, Germany}
\address[NRAL]{NRAL, Jodrell Bank, Macclesfield, Cheshire, SK11 9DL, UK}


\begin{abstract} 
  
  We discuss the technique of Wide-field imaging as it applies to Very
  Long Baseline Interferometry (VLBI). In the past VLBI data sets were
  usually averaged so severely that the field-of-view was typically
  restricted to regions extending a few hundred milliarcseconds from
  the phase centre of the field. Recent advances in data analysis
  techniques, together with increasing data storage capabilities, and
  enhanced computer processing power, now permit VLBI images to be made
  whose angular size represents a significant fraction of an individual
  antenna's primary beam. This technique has recently been successfully
  applied to several large separation gravitational lens systems,
  compact Supernova Remnants in the starburst galaxy M82, and two faint
  radio sources located within the same VLA FIRST field. It seems
  likely that other VLBI observing programmes might benefit from this
  wide-field approach to VLBI data analysis.
  
  With the raw sensitivity of global VLBI set to improve by a factor
  4-5 over the coming few years, the number of sources that can be
  detected in a given field will rise considerably.  In addition, a
  continued progression in VLBI's ability to image relatively faint and
  extended low brightness temperature features (such as hot-spots in
  large-scale astrophysical jets) is also to be expected.  As VLBI
  sensitivity approaches the $\mu$Jy level, a wide-field approach to
  data analysis becomes inevitable.

\end{abstract} 



\end{frontmatter}

\section{Introduction}
\label{intro} 

The undistorted field of view of a given VLBI data set is usually
limited by two main effects: bandwidth smearing and time-average
smearing \citep{Bridle89}. The narrower the individual frequency
channels and the smaller the integration time, the larger the
unaberrated field of view.  Data generated by VLBI correlators are
comprised of a set of measurements of the complex visibility as a
function of frequency (or delay) and time.  Most continuum VLBI data
sets are delivered to the astronomer with relatively narrow frequency
channels ($\sim 0.5$~MHz) and short integration times ($\sim 2$~secs).
For example, a typical $\lambda 18$~cm EVN data set, in its original
form, boasts a field-of-view, $\theta_{fov}$, in excess of $\sim
\frac{1}{2}$~arcminute. The same EVN data set, having for example being
averaged in frequency over $\sim 64$ ~MHz, has a field-of-view of $
\sim 300$ milliarcseconds (mas). This reduction in the field-of-view by
a factor in excess of two orders of magnitude is often considered to be
unimportant. The aim of this paper is to show, by illustration, that
this presumption may no longer be valid.

\begin{figure}[hbt] 
\centering
\includegraphics[scale=0.8,angle=0]{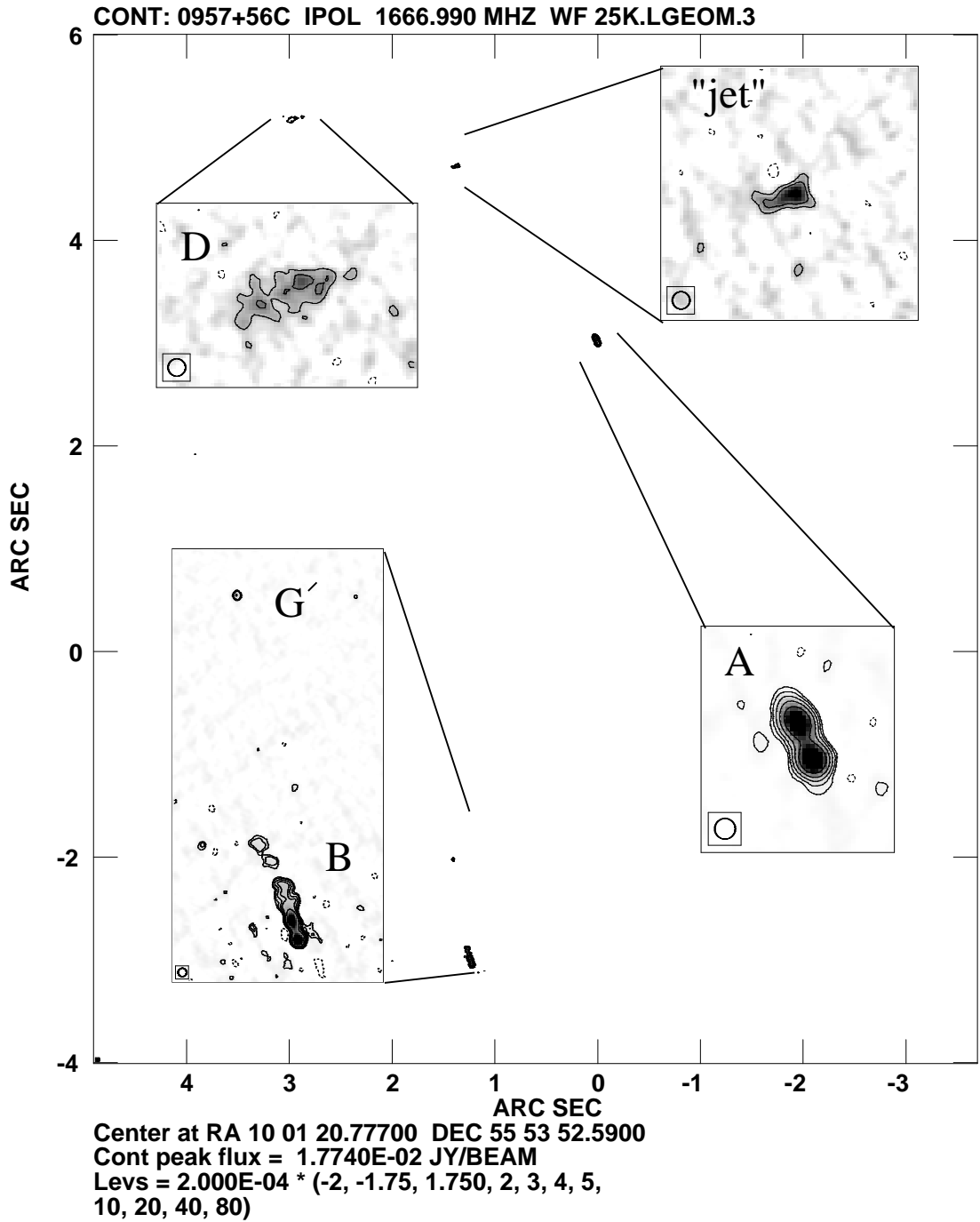}  
\caption{A wide-field image of the gravitational lens system 0957+561. 
In addition, to the previously known compact features (A,B, G$^{\prime}$)
we also detect compact structure in the arcsec-scale jet (previously 
labelled in VLA maps of the source as features ``jet'' and ``C''.} 
\label{fig1} 
\end{figure}

\section{New Developments in Wide-Field VLBI Imaging} 

\label{new_world} 
 
In the early 1980s, the off-line computer resources available to most
astronomers were ill-equipped to deal with extremely large and
cumbersome VLBI data sets. Not surprisingly, one of the major goals of
VLBI {\it data reduction} was to severely average continuum data at the
earliest possible stage in the analysis process (as soon as
fringe-fitting corrections had been applied). Today, the processing
power and data storage volumes enjoyed by the vast majority of VLBI
astronomers is $\sim 2$ orders of magnitude greater than the shared
systems used previously. Nevertheless, the custom of excessive data
averaging continues. This practice severely limits the natural field of
view of VLBI images, and is often inappropriate - especially in the era
of high sensitivity observations.

\subsection{Detection of hot-spots in Large-scale Jets }  
\label{WF_HS}

In Fig. \ref{fig1} we present a $\lambda = 18$~cm EVN-only map of the
gravitational lens, 0957+561~A,B. The data were calibrated in the usual
manner. To avoid bandwidth smearing, the data were held in the form of
28 independent but contiguous 2 MHz channels. The data were averaged in
time in a baseline dependent manner with integration times ranging from
2.5 seconds (on the longest projected baselines) to 30 seconds (on the
shortest projected baselines). Fig. \ref{fig1} clearly shows the two
main, compact components, 0957+561~A,B and a very faint compact source
known as G$^{\prime}$, lying about 1 arcsec to the north of B. All
three components have been detected by previous VLBI campaigns.
However, the real excitement relates to the fact that we have detected
and imaged two low brightness temperature features ($T_{b} \sim
10^{6}-10^{7}$ K), that are associated with compact regions or
hot-spots in the singly imaged arc-second scale jet that dominates VLA
maps of this source.  There must be many more cases where similar low
surface brightness emission goes undetected, simply because a
wide-field approach to the data analysis is not pursued.  VLBI
observations of such emission could improve our understanding of
large-scale jet physics, distinguishing between various hot-spot models
and allowing a comparison between the properties of the jet (e.g. flow
velocity) and the intergalactic medium on pc and kpc scales.


\subsection{Imaging Faint SNR in the Starburst Galaxy M82} 
\label{WF_M82} 

\begin{figure}[htb] 
\centering
\includegraphics[scale=0.5,angle=-90]{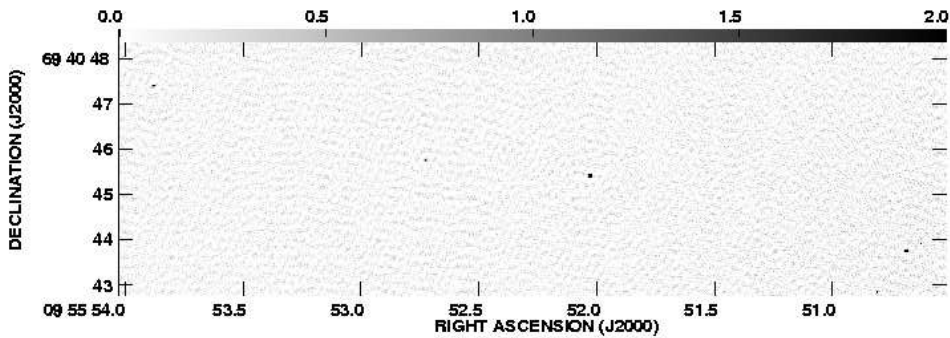}  
\caption{A large wide-field image of part of the M82 field 
imaged by \citet{Pedlar99}. Four supernovae are clearly detected in this
image, only the brightest (bottom right hand corner) was detected by
previous VLBI observations.}
\label{fig2} 
\end{figure}

A wide-field approach has also been applied by \citet{Pedlar99} to
$\lambda 18$~cm EVN observations of the starburst galaxy M82.  Previous
VLBI observations had focussed on the brightest SNR (41.95+575), but by
following a wide-field approach to the data analysis \citet{Pedlar99}
have been able to generate exquisite images of 4 other compact SNR in
the field, of which the faintest has a peak flux of $\sim
0.4$~mJy/beam. A sub-section of the entire 1 arcminute field is shown
in Fig.~\ref{fig2}. \citet{Pedlar99} have also re-analysed ``vintage''
EVN data of M82 from epoch 1986.  By employing a wide-field approach to
the re-analysis they have been able to measure expansion rates for one
of the SNR and place upper-limits on two others. This work is a superb
example of the type of information ``gain'' one can easily acquire
through the general application of wide-field techniques to VLBI data.

\subsection{Going Deeper into a Crowded Sky} 
\label{features} 

\begin{figure}[hbt] 
\centering
\includegraphics[scale=0.6,angle=-90]{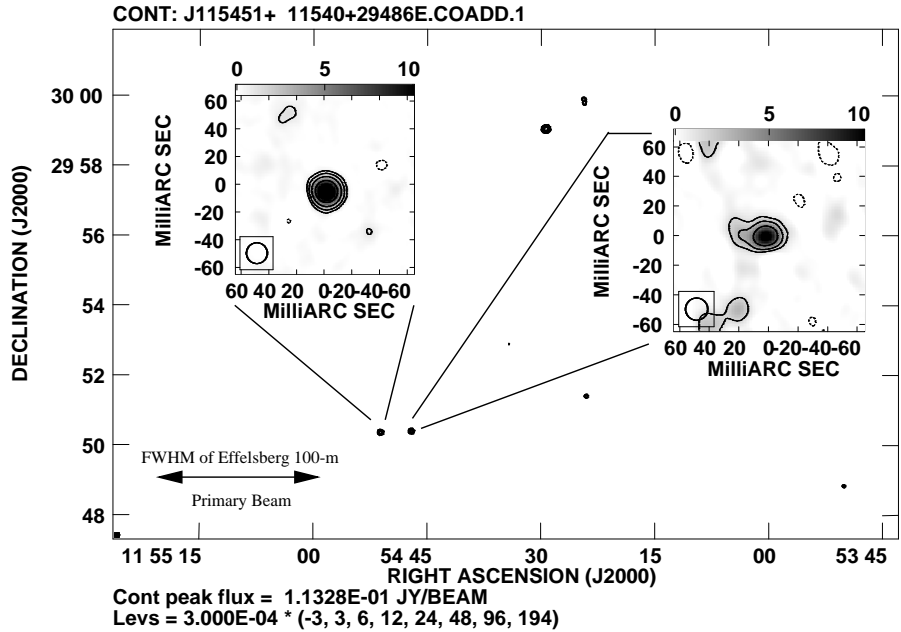}  
\caption{Global $\lambda 6$~cm VLBI maps of two closely separated 
  VLA FIRST radio sources generated simultaneously using wide-field
  data analysis techniques. Both sources are members of a faint source
  survey (see \citet{Garrington99}).  The total number of sources in
  this VLA FIRST field gives a good impression of how often one can
  encounter secondary sources at the mJy level. As VLBI sensitivity
  approaches the $\mu$Jy level, a wide-field approach to data analysis
  is inevitable. }
\label{fig3} 
\end{figure}

The advances in VLBI hardware described elsewhere in this volume
suggest that for a Global VLBI array r.m.s. noise levels of $\sim
10\,\,\mu$Jy/beam will be attainable by the end of the millennium.
Looking further ahead (10-15 years) there is every reason to believe
that $\sim 1\,\,\mu$Jy/beam noise levels will be achievable.  At this
level of sensitivity, the radio sky becomes a very crowded place. At
the $\mu$Jy level one may expect to encounter 1 source every few
arcseconds (see \citet{Muxlow99}). We can expect to reach these levels
of sensitivity within the next 10-20 years, by which time wide-field
VLBI imaging may have evolved into a standard VLBI processing route in
order to avoid source confusion. Even at the level of a few mJy a
wide-field approach may pay dividends.  \citet{Garrington99} have
embarked on a survey of faint mJy sources located within a few degrees
of a bright, compact radio source, 1156+295. Two of the faint sources
surveyed are separated by only a few arcminutes on the sky.  Even
although the observations had not been set up with wide-field imaging
specifically in mind, it was still possible to produce tapered images
of both sources simultaneously from a single $\lambda 6$~cm global VLBI
data set (see Fig. \ref{fig3}). We note that this wide-field approach
permits a very {\it direct} measurement of the relative
astrometric positions of radio sources in the same undistorted field of
view.

\section{Current Limitations} 
\label{limits} 

Currently the main limits on wide-field imaging are set by the maximum
data rate that can be generated by VLBI correlators. In other words, the
limitation is exactly the reverse of the situation in the early
1980's, where the bottleneck was associated with off-line data
processing facilities. Another restriction is the size of the primary
beam of the larger VLBI antennas, especially phased arrays (e.g. WSRT
\& VLA$_{27}$). In this case at least, ``small is beautiful''.  This
latter point should be borne in mind when the next generation of large
telescope arrays are being designed (e.g. SKA), especially with regard
to their possible incorporation within existing VLBI networks.

%







\end{document}